\documentclass[]{article}
\usepackage[caption=false]{subfig}
\usepackage{amsmath,amssymb,bm,graphicx}
\usepackage{amsthm}
\usepackage{color,grffile}
\usepackage{times}
\usepackage{rotating}
\usepackage{bm}
\usepackage[plain,noend]{algorithm2e}
\usepackage{dsfont}

\usepackage[numbers,sort&compress]{natbib}
\bibpunct[, ]{[}{]}{,}{n}{,}{,}
\makeatletter
\def\NAT@def@citea{\def\@citea{\NAT@separator}}
\makeatother

\newcommand{\Nbsim}{{T=100}}
\newcommand{\Nbvar}{{p}}
\newcommand{\nbvar}{{j}}
\newcommand{\Nbind}{{n}}

\newcommand{\Nbvardraw}{{k}}
\newcommand{\Nbiter}{{B}}
\newcommand{\Nbfold}{{L}}
\newcommand{\nbfold}{{l}}
\newcommand{\bfX}{{\bold{X}}}
\newcommand{\bfY}{{\bold{Y}}}

\newcommand{\bfbeta}{{\bm{\beta}}}

\title{An ensemble learning method for variable selection: application to high-dimensional data and missing values}
\author{Avner Bar-Hen\footnote{avner@cnam.fr}, Vincent Audigier\footnote{vincent.audigier@cnam.fr}\\ {\small CNAM, Laboratoire Cedric-MSDMA, 2 rue Cont\'e, 75003 Paris, France}
}

\begin{document}
\maketitle

\textbf{Abstract}
Standard approaches for variable selection in linear models are not tailored to deal properly with high-dimensional and incomplete data. Currently, methods dedicated to high-dimensional data handle missing values by ad-hoc strategies, like complete case analysis or single imputation, while methods dedicated to missing values, mainly based on multiple imputation, do not discuss the imputation method to use with high-dimensional data. Consequently, both approaches appear to be limited for many modern applications.

With inspiration from ensemble methods, a new variable selection method is proposed. It extends classical variable selection methods in the case of high-dimensional data with or without missing data. Theoretical properties are studied and the practical interest is demonstrated through a simulation study, as well as through an application to models specification in sequential multiple imputation.

In the low dimensional case, the procedure improves the control of the error risks, especially type I error, even without missing values for stepwise, lasso or knockoff methods. With missing values, the method performs better than reference selection methods based on multiple imputation. Similar performances are obtained in the high-dimensional case with or without missing values.
\vspace{1cm}

\textbf{Keywords}: Ensemble method; High-dimensional Data; Linear Regression; Missing Data; Variable Selection

\section{Introduction}
Large scale data is challenging for data visualisation, data understanding, large measurement and storage requirements, training and utilisation times, or prediction. Variable selection, such as stepwise e.g., is one of the most common strategies to tackle the issue. Many procedures of variable selection are still proposed in the modern literature such as  Lasso \cite{Tibshirani96}, Bolasso \cite{Bach08}, knockoff \cite{Barber15,Barber16} among others (see for example~\cite{Harrell15} for a review).

In this article we focus on a classical linear model framework in which a Gaussian response $Y$ is related to variables among a set of explanatory Gaussian variables $X_\nbvar$ ($\nbvar=1,\ldots,\Nbvar$). In this context, variable selection consists in identifying explanatory variables which are significantly related to $Y$. Extension to generalized linear models will be discussed.

The selection of variables is known to be a complicated problem. Among the difficulties often encountered we can cite the stability of the selected subset of variables, high dimensionality (when the number of variables is larger than the number of observations) or missing data for example.

Ensemble learning methods provide a way to improve stability. The  “stability”  of  a  variable  selection  algorithm  refers  to  the robustness of its selected variables,  with respect to data sampling and to its stochastic nature \cite{Breiman96,Meinshausen10}. Such methods consist in perturbing the data several times, applying the selection procedure on the perturbed data, and then, aggregating over all obtained subsets. For example, ensemble methods have been suggested for variable selection by random forests \cite{Genuer10} or lasso \cite{Meinshausen10}. As regards the high-dimensionality, it can be tackled by techniques like shrinkage methods (e.g. ridge regression or lasso \cite{Tibshirani96}), or by using preliminary screening steps \cite{Barber16,Wasserman09}. As regards the missing data issue, multiple imputation \cite{Schafer97,Rubin87,Little02} appears the most intensively investigated. In particular, many methods have been proposed to pool several subsets of variables obtained from each imputed data set, independently to the way used to fill-in the data \cite{Zhao17,Zahid19bis}. Other authors investigate the way to impute high-dimensional data, but independently to the variable selection procedure. Among them, \cite{Zahid19} investigate sequential imputation based on ridge regression. Note that the approach requires a screening step to deal with data set where the number of variables exceeds the number of individuals. In another spirit, \cite{Audigier16} investigate dimensionality reduction methods by proposing multiple imputation by principal component analysis. \cite{austin08} shows that the bootstrap did not improve upon traditional backward stepdown variable selection. Both methods fail at identifying the ``correct'' variables  when $p<N$.

However, in practice, we potentially face to all challenges simultaneously, making difficult to perform variable selection in a suitable way. In this paper, we propose an original variable selection method based on an ensemble learning method allowing variable selection in various cases, notably for high-dimensional data and/or missing data, while improving stability of the selection.
To achieve this goal, the main idea is to perform variable selection on random subsets of variables and, then, to combine them to recover which variables $X_\nbvar$ are related to the response $Y$. Note that ensemble learning methods for variable selection generally resample the individuals, but here, only variables are resampled. Performing variable selection on several subsets of variables solve the high-dimensional issue and allows treatment of missing values by classical techniques. More precisely, the outline of the algorithm are as follows: let consider a random subset of size $\Nbvardraw$ among $\Nbvar$ variables. By choosing $\Nbvardraw$ small, this subset is low dimensional, allowing efficient treatment of missing values by standard imputation method. Then, any selection variable scheme can be applied. We will focus on standard variable selection methods, such as stepwise, lasso, but also on a more recent method, named knockoff \cite{Barber16}, which has the specific property to be consistent. By resampling $\Nbiter$ times, a sample of size $\Nbvardraw$ among the $\Nbvar$ variables, we may count how many times, a variable is considered as significantly related to the response variable $Y$ and how many times it is not. We need to define a threshold to conclude if a given variable is significantly related to the response $Y$.
Our algorithm has a flavour of random forests  (\cite{breiman2001random}) but our aim is to find the relevant variables and not to predict the response variable $Y$. The question of missing values is generally addressed through surrogate variables and therefore not adapted for finding relevant variables of the model. In the same spirit, we can think to random subspace (see \cite{bryll2003attribute,ho1998random} among others) or random  projection (\cite{cuesta2007random}): these wrapper method are well designed for high-dimensional data but does not handle the question of missing values. Moreover, very few is known about mathematical properties (for random subspace) and the random sampling of the variable implies that some variables are more drawn than others. Testing procedure is not based on the same number of replications for each variable and this create bias. Our proposal does not suffer from these drawbacks.

In the next section, we fully describe the proposed algorithm. Rules to tune its parameters are given and mathematically justified. We also derive some theoretical properties of the algorithm. In Section~\ref{simulations}, we illustrate the relevance of the selection of variable method through a simulation study and we provide an application to models specification in sequential multiple imputation in Section \ref{Application}. Finally, a discussion about extensions closes the paper.

\section{Algorithm}
\subsection{Notation and context}
Let consider a classical linear regression model
\begin{eqnarray}
Y=X\bfbeta+\varepsilon\label{modelreg}
\end{eqnarray}
where $X=\left(X_1,\ldots,X_\Nbvar\right)$ denotes a set of $\Nbvar$ explanatory variables, $\bfbeta=\left(\beta_1,\ldots,\beta_\Nbvar\right)$ denotes the vector of regression coefficients, $\varepsilon$ is a Gaussian noise with variance $\sigma^2$ and null expectation, $Y$ is the response variable. $\Nbind$ independent realisations of $\left(Y,X\right)$ are observed, leading to a data set with $\Nbind$ rows and $\Nbvar+1$ columns.

We assume that missing values occur on covariates only, without loss of generalities \cite{Little02}.
We note $R=\left(R_1,\ldots,R_p\right)$ the missing data mechanism so that $R_\nbvar=1$ indicates variable $X_j$ is missing, and $R_\nbvar=0$ indicates variable is observed. The $\Nbind$ realisations of $R$ are assumed to be independent. We do not put any restrictions on the missing data mechanism, and any restrictions on the number of missing values in order to cover a large range of situations.

We intended to select the ``best'' subset of predictors, i.e. the subset of non-null coefficients of $\bfbeta$. 
The central premise is that the data contains many features that are either redundant or irrelevant, and can thus be removed without incurring much loss of information. Successful procedures are characterized by high predictive accuracy, yielding interpretable models while retaining computational efficiency. Penalized methods that perform coefficient shrinkage (such as lasso) have been shown to be successful in many cases. Models with correlated predictors are particularly challenging to tackle and missing data are difficult to handle \cite{Stadler12,Loh11}. Some alternative such as knockoff also provide statistical guarantees \cite{Barber15,Barber16} but have not been adapted to handle missing data. Stepwise regression is also very popular process of building a model by successively adding or removing variables based solely on the statistics such as AIC criterion or $t-$test of their estimated coefficients. Unfortunately, the model is fit using unconstrained least squares, therefore nothing can be said about the mathematical properties of the results. Furthermore, stepwise cannot be directly applied when $p>N$ or data with missing values.


\subsection{The algorithm\label{section:algo}}
As for ensemble methods, our algorithm has two steps: one which creates many regression instances and one which aggregates instances into an overall regression. More precisely, each regression instance allows to test if the relationships between (part of) explanatory variables and the response variable is significant or not. Then, we aggregate tests of the instances to obtain a global test for each variable.

To create regression instance, we sample $\Nbvardraw$ variables among the $\Nbvar$ variables. Next, a variable selection procedure is applied on the $\Nbvardraw$ variables. If the method does not handle high-dimensional data, $\Nbvardraw$ is chosen less than $\Nbind$, so that the high-dimensional issue is tackled. If the dataset has missing values, two cases can be considered: the first one is the number of individuals with missing is very small. 
 For such a case, complete-case analysis can be a sufficient strategy to solve the missing data issue. Otherwise, imputation according to a Gaussian model can be used. Multiple imputation is commonly advocated for inference in linear model. Indeed, compared to single imputation it provides unbiased point estimates as well as unbiased standard errors, although single imputation provides only unbiased point estimates. In our framework, we only aim to identify the set of variables related to the response variable and no to build confidence intervals. In this context, the benefit to use multiple imputation is not clear, while the computational benefit to use single imputation is. For this reason, we use single stochastic imputation by the multivariate Gaussian model to deal with missing values. Note that imputation methods need accounting for the type of the missing data mechanism \cite{Rubin87,Schafer97,Little02}. We will consider a classical method dealing with missing at random (MAR) mechanisms \cite{Schafer97}, but methods dedicated to missing not at random (MNAR) mechanisms could also be used.

After that, any variable selection procedure can be applied, leading to the regression instances among the $\Nbvardraw$ variables that are significantly related to $Y$ (according to a given threshold). We iterate the process $\Nbiter$ times, leading to $\Nbiter$ regression instances.


As a second step of the algorithm, the regression instances are aggregated. For each variable $X_\nbvar$, we count the ratio $r_\nbvar$ between (i) the number of times the variable $X_\nbvar$ is selected as significantly related to the response variable $Y$ and (ii) the number of times the variable is present in the subsets. We conclude that a variable $X_\nbvar$ is significantly related to $Y$ if $r_\nbvar$ is greater that a threshold $r$.

Sampling a subset of $k$ variables implies that each variable is chosen a random number of times. Therefore, a direct sampling of the variables implies that variance of $X_\nbvar$ is not constant among the variables. This behaviour is irrelevant since it required more iterations to bound the variance of $r_\nbvar$. Note that  the sampling process of a given variable can be viewed as a Bernoulli distribution and a bound on the minimum of Bernoulli can be easily obtained through Chernoff inequality. An alternative to this sampling scheme is random partitioning of the variables. If $\Nbvar$ is a multiple of $\Nbvardraw$, we have $\Nbvar/\Nbvardraw$ subsets of variables by partition. Variable selection is applied to each of the $\Nbvar/\Nbvardraw$ subsets so that $X_\nbvar$ is observed with the same proportion $\Nbvardraw/\Nbvar$ over all subsets. We iterate the process by choosing random partitions.

Three questions arise: (i) how to choose $\Nbiter$, (ii) how to choose the threshold $r$ and finally (iii) how to choose $\Nbvardraw$.

\subsubsection{How many iterations?}
To improve the stability of the procedure, the proportion of times that a variable is considered as significant ($r_\nbvar$) needs to be calculated from many iterations ($\Nbiter$). 

%
%
$\Nbiter$ has to be chosen, so that $\mathbb{V}\left(r_\nbvar\right)$ is small. If the number of times the variable $X_j$ is significant follows a Binomial distribution, then for $\tilde{\Nbiter}$ regression instances gathering $X_\nbvar$ ($\tilde{\Nbiter}=\Nbiter\times\Nbvardraw/\Nbvar $), we have $\mathbb{V}\left(r_\nbvar\right)$ is less than $\frac{1}{4\tilde{\Nbiter}}$. Thus, $\tilde{\Nbiter}=100$ can be chosen to obtain a standard error less than 5\%. For a given value of $\Nbvardraw$, it provides a guideline to tune $\Nbiter$.

We can note that $\Nbiter$ is related to $\Nbvar$, meaning that the number of iterations of the algorithm needs to be chosen according to the number of variables.

\subsubsection{Which value for the threshold $r$?}
\label{section_thres}
 $r_\nbvar$ can be seen as a variable importance measure or more precisely as an estimate of  $\alpha$, the risk of the test between the null hypothesis $H_0: ``\beta_\nbvar= 0`''$ versus the alternative $H_1: ``\beta_\nbvar\neq 0''$ over all the $\Nbiter$ iterations. Following  Neyman-Pearson lemma, $r_\nbvar$ needs to be chosen over than a threshold $r$ (e.g. 95\%) to control the $\alpha$ risk.

To tune the threshold $r$ whatever the data set, theoretical properties of the selection method are needed. For instance, the false discovery rate is controlled by knockoff at each iteration. Thus, it could be preserved by choosing $r$ accordingly.
However, for many variables selection methods no such guaranties are available. For them, $r$ can be chosen a posteriori by empirical methods like cross-validation.
For achieving this goal, the algorithm is applied on a train set. From the variable importance of each variable, a sequence of nested linear regression models can be derived. For each one, an error of prediction can be calculated and the threshold is chosen as the one minimising this error.  More precisely, we proceed by $k$-fold cross validation as follows:
\begin{enumerate}
\item Split the data $\left(\bfX,\bfY\right)$ in $\Nbfold$ folds $\left\lbrace\left(\bfX^1,\bfY^1\right),\left(\bfX^2,\bfY^2\right),..,\left(\bfX^\nbfold,\bfY^\nbfold\right),...,\left(\bfX^\Nbfold,\bfY^\Nbfold\right)\right\rbrace$ of size $\left(\lfloor\Nbind/\Nbfold\rfloor \times \Nbvar +1\right)$
\item For $\nbfold\in\{1,...,\Nbfold\}$
\begin{enumerate}
\item Based on $\left(\bfX^\nbfold,\bfY^\nbfold\right)$, compute the variable importance of each variable $\left(r_\nbvar^\nbfold\right)_{1\leq\nbvar\leq\Nbvar}$ using the proposed algorithm
\item Order variables by decreasing importance. The set of ordered variables is denoted $\left(X_{1^\star},...,X_{\Nbvar^\star}\right)$ and the set of ordered variable importance is denoted $\left(r_{\nbvar^\star}^\nbfold\right)_{1\leq\nbvar^\star\leq\Nbvar}$
 \item Associate a sequence of nested models $\mathcal{M}^\nbfold_1,\mathcal{M}^\nbfold_2,...,\mathcal{M}^\nbfold_\Nbvar$ so that $\mathcal{M}^\nbfold_\nbvar$ $(1\leq\nbvar\leq\Nbvar)$ is the linear regression model where $Y$ is explained by $\left(X_{1^\star},...,X_{\nbvar^\star} \right)$
\item \label{steppb} For all $1\leq\nbvar\leq\Nbvar$, predict $\bfY^\nbfold$ according to $\mathcal{M}^\nbfold_\nbvar$ and compute the prediction error $PE^\nbfold_\nbvar$
\end{enumerate}
\item Choose $r$ according to
\begin{equation}
\operatorname*{argmin}_{r\in]0;1]} \sum_{\nbfold=1}^{\Nbfold} \sum_{\nbvar=1}^\Nbvar \mathds{1}_{]r_{\nbvar^\star-1}^\nbfold;r_{\nbvar^\star}^\nbfold]}(r)PE_\nbvar^\nbfold
\end{equation}
\end{enumerate}

Note that in a context of regression, optimising $r$ in terms of prediction error is equivalent to optimisation in terms of regression coefficient estimate, making cross-validation consistent with identification of non-null regression coefficients.

High-dimensional data or missing values are tricky for cross-validation, since linear models cannot be directly fit in  step \ref{steppb} in both cases. The high-dimensional issue can be tackled by only considering models $\mathcal{M}^\nbfold_1,\mathcal{M}^\nbfold_2,...,\mathcal{M}^\nbfold_\Nbind$ (since their number of explanatory variables is less than $\Nbind$), while the missing data issue can be handled by imputing the test set and train set simultaneously, excluding the response variable on the test set as proposed in \cite{Kapelner15}.


%
\subsubsection{What is the optimal size of regression instances $k$?}
\label{sectionbruit}
The practical usefulness to perform variable selection from a subset of $\Nbvardraw$ variables instead of $\Nbvar$ have been already explained at the beginning of Section \ref{section:algo}. We now highlight how does this strategy influence the performances of a selection procedure in the case of complete data.\\

 For a regression instance, let's define $\delta_\nbvar$ such that  $\delta_\nbvar=1$ if $X_\nbvar$ is drawn and zero otherwise.
Putting them in a diagonal matrix $\Delta=diag(\delta_1,\ldots,\delta_p)$, the regression model based on a sample of $\Nbvardraw$ variables can be written as:
\begin{equation*}
  Y= X\Delta\beta+\varepsilon
\end{equation*}
where $X\Delta$ corresponds to the design matrix constructed on selected variables

Without loss of generality, consider that $X$ gathers significant explanatory variables only (i.e. $\beta_\nbvar\neq 0$ for all $1\leq \nbvar\leq \Nbvar$). Then, by independence between $\varepsilon$ and $X$ \begin{eqnarray}
  \mathbb{V}(Y\vert X\Delta)&=&\mathbb{V}(X(I-\Delta)\beta)+\mathbb{V}(\varepsilon)\\
  &=&\mathbb{V}(X(I-\Delta)\beta)+\mathbb{V}(Y\vert X).
\end{eqnarray}

The higher the proportion of significant variables not present in the regression instances, the larger is $\mathbb{V}(X(I-\Delta)\beta)$   (and $\mathbb{V}(Y\vert X\Delta)$ \textit{a fortiori}). This implies that the regression scheme will be noised if relevant significant variables are missed. This situation arises when the variables are sampled through the algorithm, but identifying significant variables is more challenging on noisy data. Thus, to limit this loss of power, it seems more relevant to consider a large value for the number of selected instances ($\Nbvardraw$). In practice, we propose to consider $k$ ten times smaller than the number of observations. This rule of thumb is robust to the choice of $k$ since the threshold $r$ compensates its misspecification. Based on simulations, this will be discussed in the Section \ref{simulations}.


These results imply relationships between $\Nbvardraw$ and $r$. More precisely if $\Nbvardraw$ is small, error of the model often included significant variables and the ratio signal/error is lower. Consequently, identifying significant variables is more challenging and the proportion of times a variable is considered as significant will be smaller. To account this, the threshold $r$ should be also smaller. Thus, the choice of $\Nbvardraw$ is essentially tuned by tuning $r$.

The sensibility of $r$ to $\Nbvardraw$ will be discussed in the simulations presented in the Section \ref{simulations}.


\subsection{About aggregation of regression coefficients}
\label{math_properties}
\label{math}

Even if our goal is only to identify the subset of variables related to the response $Y$, we briefly investigate the performances of the aggregation of the regression coefficients estimates obtained by averaging of the $\Nbiter$ instances.\\

At first, let consider the sampling of variables for a regression instance and assume for the moment the absence of missing data. The regression model (Equation \ref{modelreg}) based on a sample of $\Nbvardraw$ variables can be rewritten as:
\begin{eqnarray*}
  Y&=&X\beta+\varepsilon\\
  &=& X\Delta\beta+X(I-\Delta)\beta+\varepsilon\\
  &=&X\Delta\beta+\varepsilon'
\end{eqnarray*}
where $\varepsilon'\sim{\cal N}(X(I-\Delta)\beta ,\sigma^2I)$. Since $\Delta$ is a projection matrix, then $\Delta^2=\Delta$ and $(X\Delta)(\Delta\beta)=X\Delta\beta$.

We assume that $X$ is invertible and, by convention, $0/0=0$. If $X$ is not invertible, we consider generalized inverse as Moore-Penrose pseudo inverse
$$\Delta\hat\beta=(\Delta X'X\Delta)^{-1}\Delta X'Y$$
and
\begin{eqnarray}
  \mathbb{E}(\Delta\hat\beta)&=&(\Delta X'X\Delta)^{-1}\Delta X'\mathbb{E}(Y)\nonumber\\
&=& (\Delta X'X\Delta)^{-1}\Delta X'(X\Delta\beta+X(I-\Delta)\beta)\nonumber\\
  &=&\Delta\beta+(\Delta X'X\Delta)^{-1}\Delta X'X(I-\Delta)\beta\label{eq1}\\
  \mathbb{V}(\Delta\hat\beta)&=&(\Delta X'X\Delta)^{-1}\sigma^2\label{eq2}
\end{eqnarray}

We see from  Equation~(\ref{eq1}) that the bias of $\hat\beta$ is induced by the correlation between the subset of variables in the regression instance and the other variables that are not selected in the regression instance. Thus, aggregation of regression estimates by averaging is relevant if and only if the design is orthogonal and very tricky otherwise.\\

\section{Simulations}\label{simulations}
\subsection{Simulation design}
To study the quality of the procedure we simulate various cases for the number of variables ($\Nbvar$), the correlation between covariates ($\rho$), the signal to noise ratio ($snr$), the nature of the missing data mechanism. Note that the snr is defined as $\mbox{snr}=\frac{||X\beta||_2}{||e||_2}$ and measures the ratio between model and noise, \emph{i.e.} the proportion of signal within the observations. For each configuration, $\Nbsim$ data sets are generated, and for each one, variable selection is performed according to the proposed algorithm and methods presented below (Section~\ref{simu:methods}).

\subsubsection{Data generation}
For a given configuration, data sets are generated as follows. First, $\Nbind$ observations for $p$ covariates are generated according to a multivariate normal distribution with null expectation, and variance
{\tiny
$\left(\begin{array}{ccccc}
1&\rho&\ldots&\ldots&\rho\\
\rho&1&\rho&\ldots&\rho\\
\vdots&\ddots&\ddots&\ddots&\vdots\\
\rho&\ldots&\rho&1&\rho\\
\rho&\ldots&\rho&\ldots&1\\
\end{array}\right)$}
where $-1\leq\rho\leq 1$.
Then, the response $Y$ is simulated according to a linear model $Y=X\bm{\beta}+\varepsilon$ where $\varepsilon$ is Gaussian with null expectation and variance $\frac{1}{snr+1}$, $\bm{\beta}$ is the sparse vector of regression coefficients composed of zeros and a fixed value $\beta$ only. This value is chosen so that $\text{Var}\left(Y\right)=1$.

For all configurations, we keep:
\begin{itemize}
\item the number of individuals $\Nbind=200$
\item and the number of non-zero values in $\bm{\beta}$ is $8$.
\end{itemize}
Before introducing missing values, configurations vary only by the values of $\Nbvar$, $\rho$ and $snr$:
\begin{itemize}
\item we consider $\Nbvar=100$ or $\Nbvar=300$ variables. Let note that for the second case, the number of variables is higher that the number of observations
\item we test two cases for the correlation $\rho=0$ and $\rho=0.4$. Note that a correlation of 0.4 between all pairs of variables is high. High correlation among explanatory variables often generates spurious results for variable selection
\item finally, we test $snr=2$ and $snr=4$, by tuning $\beta$ and the variance of the noise under the constraint that the variance of $Y$ is equal to one. Each case corresponding to high or low difficulty to select relevant variables.
\end{itemize}


\subsubsection{Missing data mechanisms}
Next, covariates of each data set are set to missing according to several mechanisms. We consider a missing completely at random (MCAR) mechanism, so that $\mathbb{P}\left(R_j=1\right)=a$ for all $j$ ($1\leq j\leq p$) and a MAR mechanism, so that $\mathbb{P}\left(R_j=1\vert Y\right)=\Phi\left(a+Y\right)$ with $\Phi$ the cumulative distribution function of the standard normal distribution. 
The coefficient $a$ of those models is tuned to get (in expectation) 20\% of missing values. The MCAR mechanism is a particular case of MAR mechanism, which is generally simpler to handle.

\subsubsection{Methods}
\label{simu:methods}
Parameters of the algorithm are tuned as follows: at each iteration $\Nbvardraw=6$ variables are drawn when $\Nbvar=100$, while $\Nbvardraw=10$ variables are drawn when $\Nbvar=300$; $\Nbiter=6000$ iterations are performed; variables that are selected at least $r= 95\%$ of the time are kept. Sensitivity to the parameters $\Nbvardraw$, $\Nbiter$ and $r$ is assessed in Section \ref{sectioninfluencepar}.\\

The investigated variable selection procedures are the knockoff, the lasso and the stepwise (with AIC). In any configuration, these methods can be used through the proposed algorithm, but not directly on the full data set because they have some lacks with high-dimensional data and or missing values.
Thus, we make comparisons as follows: we first generate the data sets (without missing data) and apply knockoff, lasso as well as stepwise variable selection procedure. Two versions of the knockoff are available: the fixed-X knockoff and the model-X knockoff. According to recommendations \cite{knockoffpackage}, we use fixed-X knockoff for low dimensional data and model-X knockoff for high-dimensional data. Note that in the proposed algorithm, only fixed-X knockoff is used. High-dimensional setting is tackled by a screening step in stepwise.

Then, we generate the missing values according to a pre-defined missing data mechanism. If possible, knockoff, lasso and stepwise variable selection are applied using complete case analysis. Note that handling missing values by imputation would be challenging here because of the large number of variables compared to the number of individuals \cite{Audigier16}. The proposed algorithm is also applied by using knockoff, lasso and stepwise variable selection where missing values are handled by single stochastic imputation according to the Gaussian model. In addition, we make comparison with a recent method combining multiple imputation and random lasso variable selection \cite{Liu16} named MIRL. This method consists in performing multiple imputation by chained equations to fill the data, then applying random lasso on imputed data sets and combining selected subsets of variables. Since multiple imputation by chained equations is too much time consuming for large data sets, we cannot apply it for high-dimensional data.\\

All computations were performed using R \cite{R}. Lasso was computed using the library \emph{glmnet}, knockoff using the library \emph{knockoff} and stepwise using the library \textit{stats}. The R code used for MIRL has been obtained from authors. Single stochastic imputation by the Gaussian model has been performed with the library \emph{norm}. The R code used for simulations is available on demand.

\subsection{Results}

\begin{sidewaysfigure}
\centering
\subfloat[Low-dimensional setting\label{low}]{
        \resizebox*{10cm}{!}{\includegraphics[width=10cm,height=15cm]{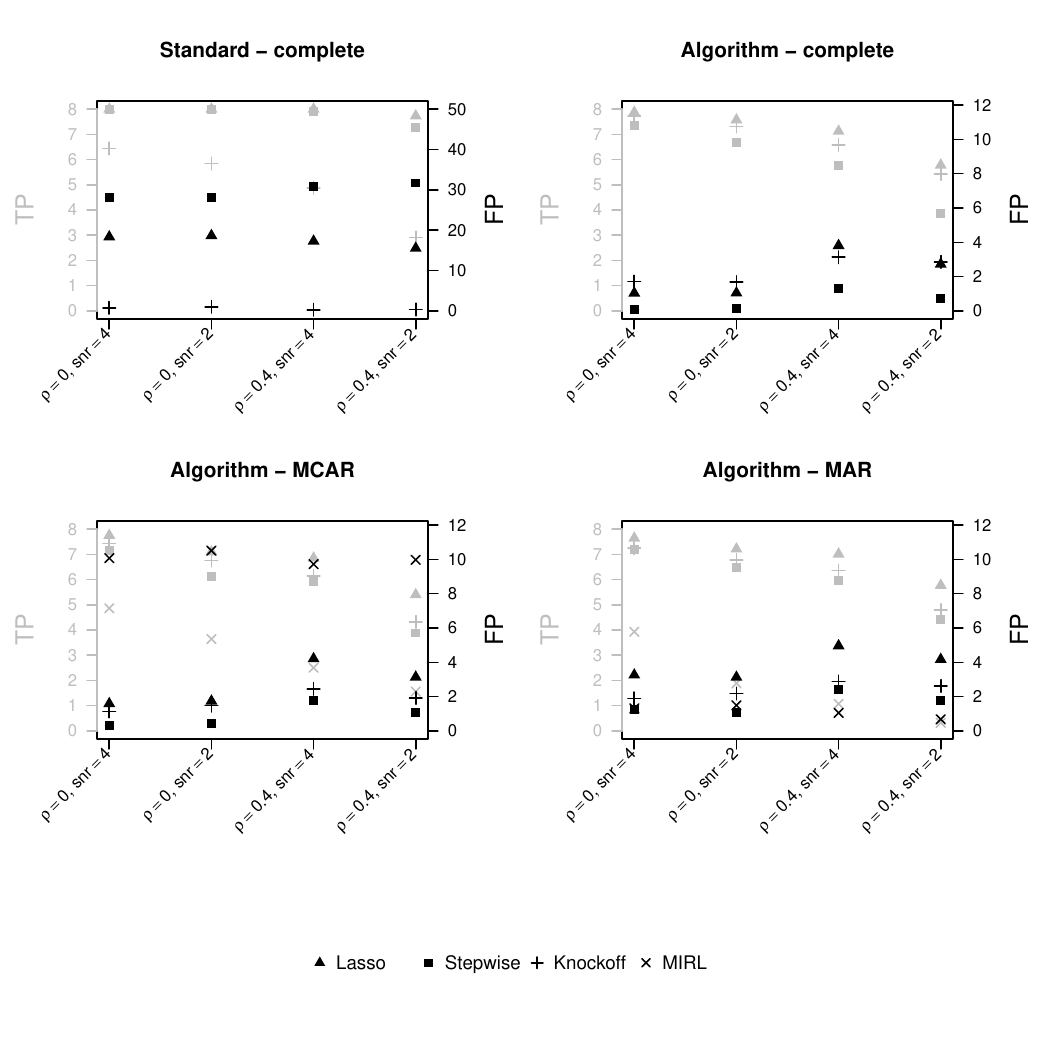}}}\subfloat[High-dimensional setting\label{high}]{
        \resizebox*{10cm}{!}{\includegraphics[width=10cm,height=15cm]{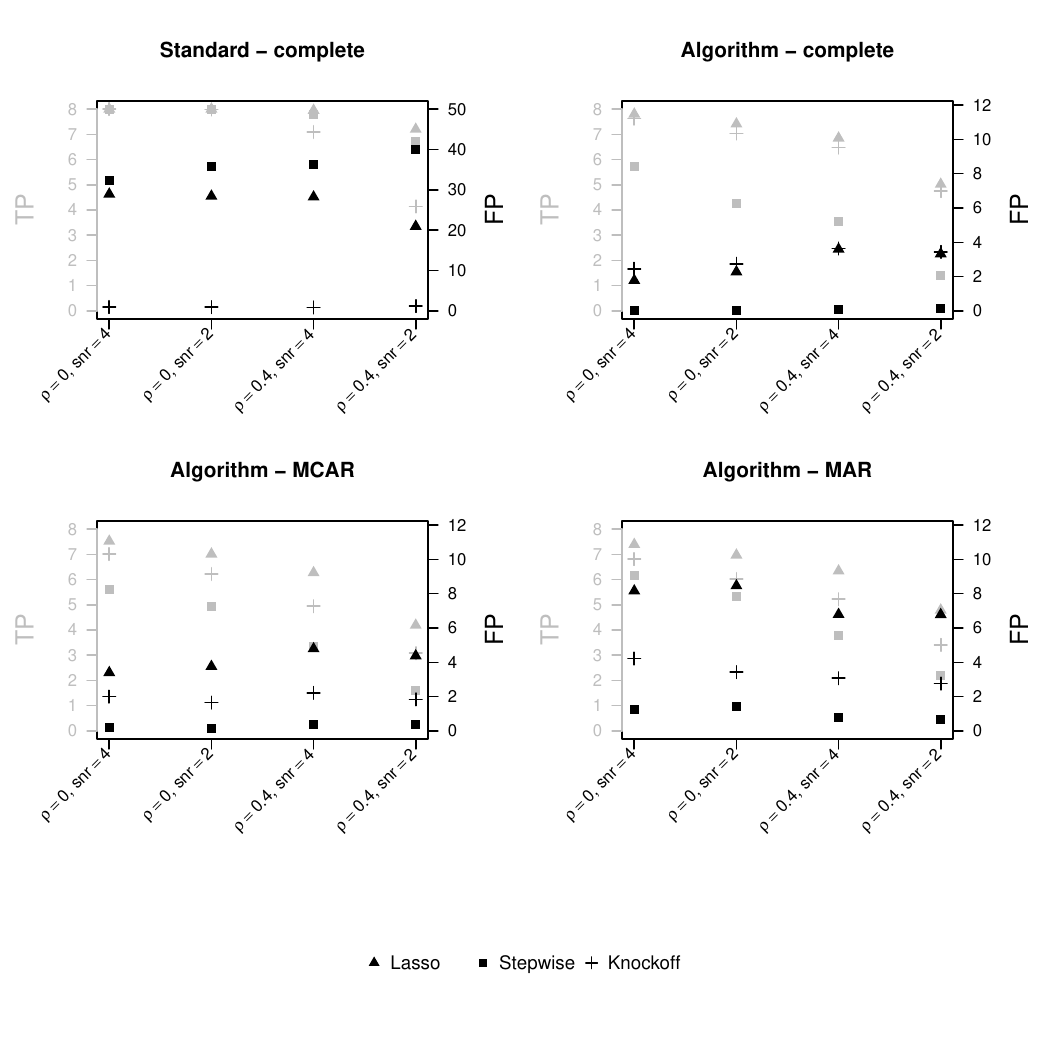}}}
\caption{Assessment of the ensemble method according to the number of variables, the missing data mechanism and the data structure. Figure \ref{low} reports results when $\Nbvar=100$ (low dimension) and Figure \ref{high} reports results when $\Nbvar=300$ (high-dimensional setting). Assessment is based on data sets varying by the correlation between covariates ($\rho$), the signal to noise ratio ($snr$) and the missing data mechanism (complete, MCAR, MAR). For a given configuration, $\Nbsim$ data sets are generated. When data are complete (see  the top of Figure \ref{low} and Figure \ref{high}), three variable selection methods (Knockoff, Lasso and Stepwise) are compared when they are directly applied on the data set (standard) or when they are iteratively applied on subsets of variables (algorithm). When data are incomplete, only the ensemble method can be applied. MIRL is also investigated when the dimension is low. Performances of the selection procedure are assessed by: the mean number of true positives (in grey) and the mean number of false positives (in black) over the $\Nbsim$ data sets.} \label{figure_main}
\end{sidewaysfigure}

We split results in four parts depending if we consider low/high dimension and complete or incomplete data (results shown in Figure \ref{figure_main}). We successively present the results for the four ones. Finally, we study the robustness of our algorithm to the tuning parameters.

\subsubsection{Low dimensional data without missing values}
In the case $\Nbind>\Nbvar$ without missing values, direct application of any standard selection variable procedures can be performed. The top-left of Figure \ref{low} reports a very large number of false positives, FP, (over than 15) for lasso and stepwise. On the contrary, this number is well controlled by knockoff (close to 1), while having many true positives, TP, (over than 5) even when the signal to noise ratio is small or correlation between covariates is large.

With our ensemble method (see  the top-right of Figure \ref{low}), the selection based on knockoff shows larger number of TP and FP than its direct application on the data set (see  the top of Figure \ref{low}). For stepwise and lasso, performances are much better improved by our algorithm whatever the correlation and the signal to noise ratio. Indeed, the number of false positives becomes close to 0.

Complementary results with 1000 and 10000 observations are reported in Figure \ref{figure_pt_dim_plusind} in Appendix. They highlight the performance of the algorithm can be deteriorated when variables are correlated and the number of observations is very large. Indeed, the threshold $r=0.95$ is motivated by the independence case, where explanatory variables are uncorrelated. However, it is only an approximation otherwise. Thus, since a large number of observations tends to increase the number of positives, the misspecification of the threshold with correlated variables tends to increase false positives.

\subsubsection{Low dimensional data with missing values}
The bottom of Figure \ref{low} and Table \ref{table_ptdim_incomplet} in Appendix report simulation results when data are missing completely at random or missing at random. In such a case, lasso, knockoff and stepwise cannot be directly applied. Therefore, complete case analysis is used. Because of the decrease of the number of individuals, selection methods have less power, leading to very poor performances (results shown in Table \ref{table_ptdim_incomplet} only). Indeed, the number of true positives is close to 0 and the number of false negatives close to 8 for each of them. When applying the MIRL method, selection is also quite bad. Indeed, the issue is that the predictive distribution of missing values is not well estimated because of the too large number of variables compared to the number of individuals \cite{Audigier16}.

On the contrary, by using our algorithm, the performances are globally similar to the case without missing values (cf bottom of Figure \ref{low}).

\subsubsection{High-dimensional data without missing values}
The top of Figure \ref{high} summarizes simulation results in the case $\Nbind<\Nbvar$ without missing values. In a similar way to the case where the dimensionality is low (at top of Figure \ref{low}), our algorithm decreases the number of false positives for selection by lasso or stepwise, but does not improve performances of the knockoff. Note that the knockoff is well suited to handle high-dimensional data when data are complete \cite{Barber16}.

\subsubsection{High-dimensional data with missing values}
Bottom of Figure \ref{high} and Table \ref{table_gddim_incomplet} report results for configurations where $n<p$ with missing values generated according to a MCAR or MAR mechanism. Because of the large number of variables, MIRL method cannot be applied since the imputation becomes too much time consuming. Direct application of variable selection methods by complete case analysis appears clearly irrelevant (see Table \ref{table_gddim_incomplet}). Indeed, the number of false negatives is close to 8, whatever the selection variable method: like in the low dimensional case, complete case analysis decreases the power of the tests and selection variables methods rarely reject the null hypothesis.

On the opposite, our algorithm leads to a small number of FP, even if this number is a little higher than in the case without missing values (see  the top-right of Figure \ref{high}).

\subsection{Influence of tuning parameters}
\label{sectioninfluencepar}
To complete this simulation study, robustness to the tuning parameters is assessed. We focus on the number of variables sampled ($\Nbvardraw$), the number of iterations ($\Nbiter$) and the threshold ($r$).

\subsubsection{Influence of $\Nbvardraw$}
\begin{figure}[h]
\begin{center}
\includegraphics[width=10cm,height=8cm]{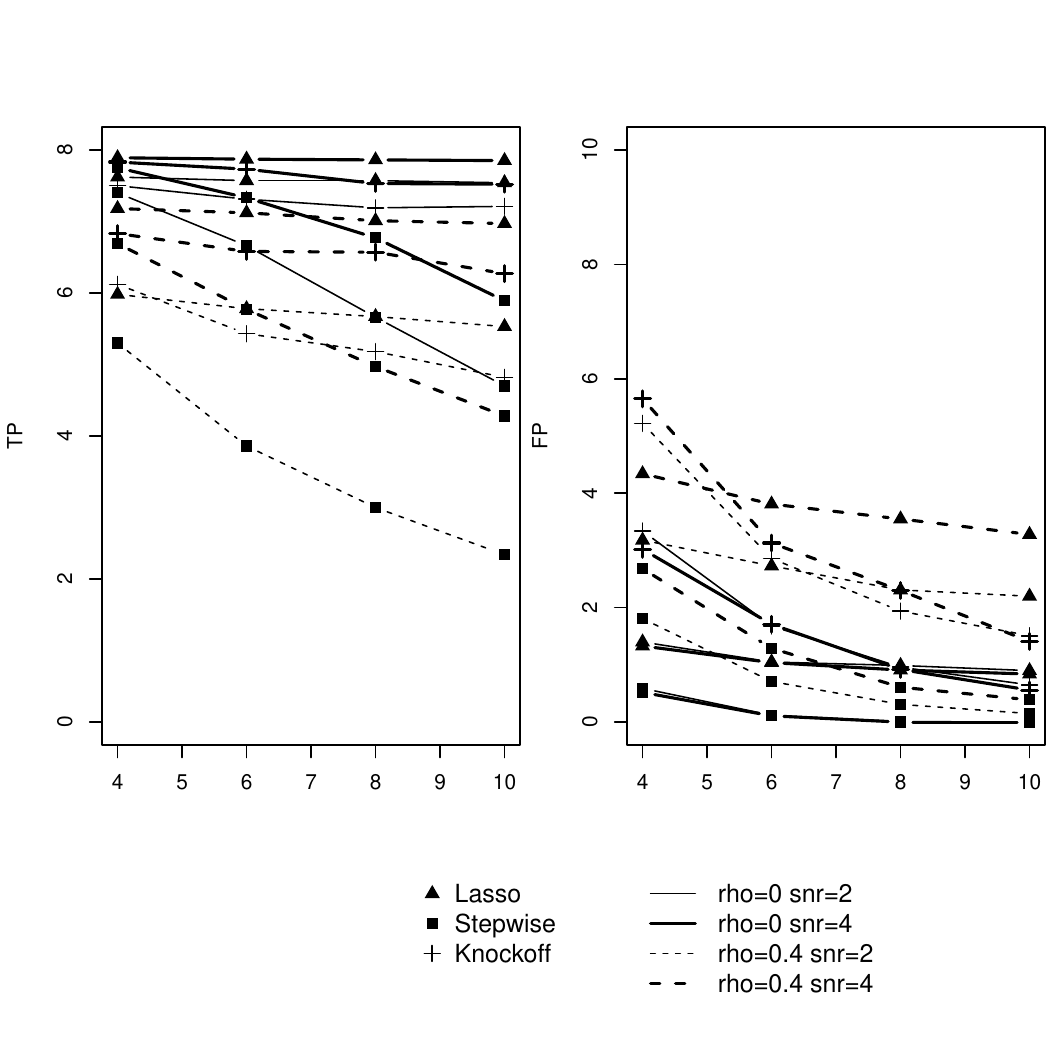}
\caption{Influence of $\Nbvardraw$ for low dimensional setting without missing values: number of true positives (on the left) and false positives (on the right) according to the number of variables sampled in the algorithm ($\Nbvardraw$) for the 4 configurations varying by the signal to noise ratio (snr) and the correlation between covariates ($\rho$). Three variable selection methods are reported (lasso, stepwise and knockoff).}\label{fig_k_complet}
\end{center}
\end{figure}

Figure \ref{fig_k_complet} reports the number of true positives and the number of false negatives according to the number of variables sampled in the algorithm (when $\Nbind>\Nbvar$ without missing values). Surprisingly, the number of true positives is globally decreasing when $\Nbvardraw$ is increasing, like the false positive rate. More precisely, the increase is important for stepwise, while it remains moderate for lasso and knockoff. The opposite could be expected since the signal to noise ratio of the regression scheme is smaller if $\Nbvardraw$ is small (cf Section \ref{math}). One possible reason is that the counterpart to increasing $\Nbvardraw$ is to decrease the degrees of freedom attributed to the model selection process. However, by drawing $\Nbvardraw$ among $\Nbvar$, the gain to increase $\Nbvardraw$ is small because the probability of selecting significant variables is small (here 8 over 300), while degrees of freedom are decreasing, implying a loss of power, but not a substantial decrease of noise on the regression scheme.

The behaviour is more severe for stepwise. Indeed, this procedure additionally often rejects the null if there are no significant variables in the subset. Such cases are more frequent when $\Nbvardraw$ is small and tends to disappear when $\Nbvardraw$ increases. 

Note that similar results are observed for the high-dimensional setting (Figure \ref{fig_k_complet_gddim} in Appendix) or when data are incomplete (Figures \ref{fig_k_incomplet_mcar} and \ref{fig_k_incomplet_mar} in Appendix).

\subsubsection{Influence of $\Nbiter$}
\begin{figure}[h]
\begin{center}
\includegraphics[width=10cm,height=8cm]{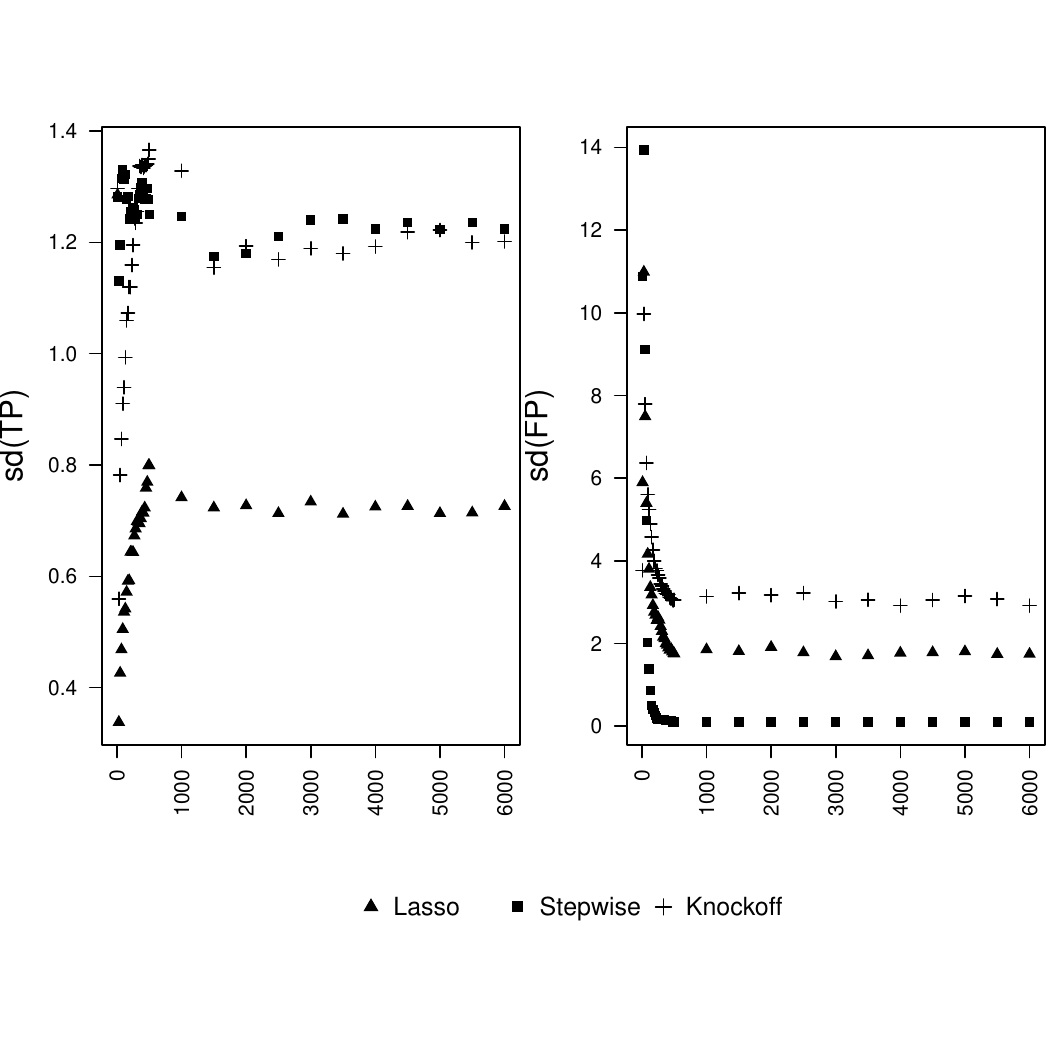}
\caption{Influence of $\Nbiter$ for high-dimensional setting without missing values: standard deviation of the number of true positives over the 100 generated data sets (on the left) and false positives (on the right) according to the number of iterations of the algorithm ($\Nbiter$) for the configuration with signal to noise ratio equal to 2 and null correlation between covariates ($\rho$). Three variable selection methods are reported (lasso, stepwise and knockoff).}\label{fig_B_complet}
\end{center}
\end{figure}

$\Nbiter$ controls the uncertainty on the proportions $\left(r_\nbvar\right)_{1\leq\nbvar\leq\Nbvar}$: for low values of $\Nbiter$, the subset of selected variables is expected to be unstable. To assess the robustness of the results to the number of iterations, we inspect the standard deviation of the number of false positives and true positives over the $\Nbsim$ generated data set according to the number of iterations. For simplicity, we only inspect 2 configurations: in the first one, data are complete with a signal to noise ratio of 2, null correlation between covariate and $\Nbind<\Nbvar$ (Figure \ref{fig_B_complet}), while data are missing according to a MAR mechanism for the second one (Figure \ref{fig_B_incomplet} in Appendix). As expected, in both cases, the variability of the TP and FP is decreasing and reaches convergence before 1000 iterations. This result is directly related to the variance of a proportion as mentioned in Section \ref{sectionbruit}. Furthermore, the number of true positives is more stable than the false negative one, which is directly related to the larger number of negatives than positives in the data. For comparison, when selection variables methods are directly applied, standard deviation for TP is 0.14 for knockoff (1.29 for FP), 0 for lasso (13.35 for FP) and 0 for stepwise (6.34 for FP).

\subsubsection{Influence of $r$}
\begin{figure}
\small
\begin{center}
%
\includegraphics[scale=.8]{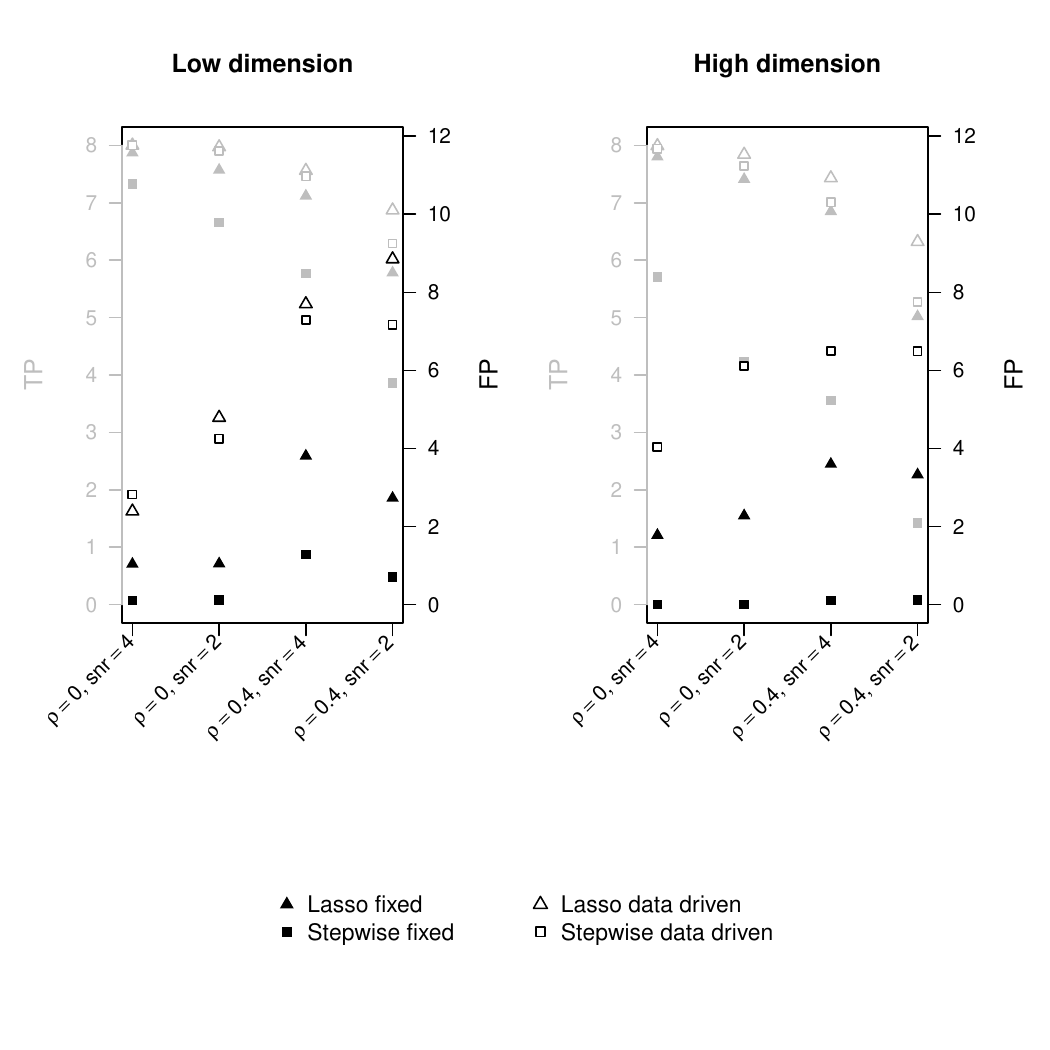}
\caption{Influence of the thresholding in the low dimensional (left) and high (right) setting without missing values: illustration for Lasso and Stepwise by tuning the threshold by cross-validation or by fixing it to 0.95. Data sets vary by the correlation between covariates ($\rho$) and the signal to noise ratio ($snr$). For a given configuration, $\Nbsim$ data sets are generated and performances of the selection procedure are assessed by: the number of true positives and the false positive (the number of real positive is 8 and the number of real negative is 92 or 292).
\label{figure_complet_cv}}
\end{center}
\end{figure}

Of course, the threshold $r$ allows a control on the FP and the TP since it is in bijection with the number of positives. According to the selection method used, $r$ can sometimes be tuned a priori, but in many cases, it should be driven by data. For achieving this goal, we use cross-validation. Figure \ref{figure_complet_cv} highlights performances of the algorithm when $r$ is data driven (by using stepwise or lasso in the complete case, in the low or high-dimensional setting). In all configurations, cross-validation leads to choose a higher value of $r$ than 0.95, which decreases the number of positives. The gain in terms of FP is more substantial than the lost in terms of TP since the number of real false positive (92 in the low dimensional setting and 292 in the high-dimensional setting) is much higher than the number of real positives (8).

\begin{figure}[h]
\begin{center}
\includegraphics[width=10cm,height=8cm]{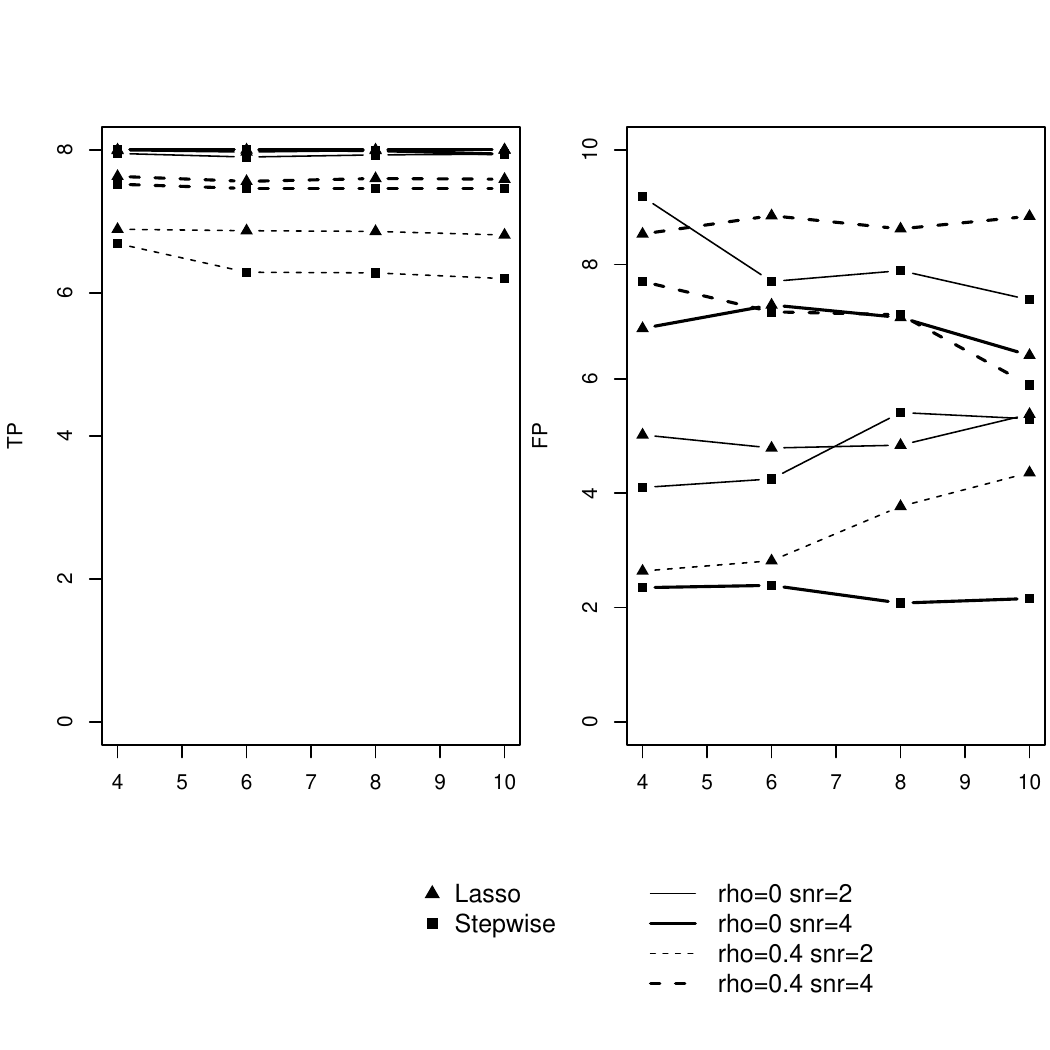}
\caption{Influence of $\Nbvardraw$ for low dimensional setting without missing values when $r$ is chosen by cross-validation: number of true positives (on the left) and false positives (on the right) according to the number of variables sampled in the algorithm ($\Nbvardraw$) for the 4 configurations varying by the signal to noise ratio (snr) and the correlation between covariates ($\rho$). Two variable selection methods are reported (lasso and stepwise).}\label{k_vc}
\end{center}
\end{figure}

Finally, we illustrate the relationship between $r$ and $\Nbvardraw$, by investigating the robustness of the procedure to $\Nbvardraw$ when $r$ is chosen by cross-validation. Results are shown in Figure \ref{k_vc}. Compared to Figure \ref{fig_k_complet} error rates are stable whatever the choice of $\Nbvardraw$. Thus, cross-validation for $r$ tuning makes the procedure robust to the choice of $\Nbvardraw$.

\subsection{Summary}
To sum up, when data are complete, the proposed algorithm significantly improves the performances (in terms of true positives and false positives) of lasso and stepwise, the gain with knockoff is less obvious. When data are incomplete, the algorithm allows to use lasso, stepwise as well as knockoff. Moreover it improves the results for these variable selection methods. As expected, its performances are globally higher when variables are uncorrelated, when the signal to noise ratio is large, or when the proportion of missing values is small (like all variables selection methods).

Concerning its tuning parameters, moderately large $\Nbiter$, say $\Nbiter=300$, can result in very stable estimate. The algorithm is robust to the choice of $\Nbvardraw$, if $r$ is well chosen. This choice can be data driven using cross-validation, but could also be tuned \textit{a priori} for knockoff. Note cross-validation remains recommended with a large number of observations and correlated variables.
\section{Application: model specification in sequential imputation\label{Application}}
As explained in the introduction, dealing with high-dimensional data is challenging for MI by chained equations. The issue can be tackled by specifying each conditional imputation model, but it remains a fastidious and highly difficult task with missing data. Thus, a natural application of the proposed algorithm is to provide an automatic choice of these models.

To illustrate this application of the algorithm, two data sets are considered: the decathlon data set and the wine data set from the R package FactoMineR \cite{Facto}. The first data set refers to $\Nbind=41$ athletes' performances during a decathlon event. It contains $\Nbvar=10$ variables corresponding to the trials. Note the number of individuals is greater than the number of variables. The second data set comes from a sensory study where $\Nbind=21$ wines of Val de Loire were evaluated on $\Nbvar=29$ descriptors. Here, the number of variables exceeds the number of individuals. Since both data sets are complete, we generate 30\% of missing values under a MCAR mechanism. 150 missing data patterns are considered. For each one, we perform multiple imputation by chained equations under linear models using $5$ imputed data sets and 10 iterations. MI is performed using the mice R package. Two options are considered: by considering all variables or by specifying conditional models. In this later case, models are specified as follows: for each incomplete variable, variable selection is performed using knockoff with a threshold $r=0.95$. The number of variables $k$ is tuned to the number of observed individuals on the variable divided by 10. $B=300$ iterations are used for the decathlon data set and $B=800$ for the wine data set. Note that the mice R package automatically performs a quick variable selection if the imputation models cannot be fit (see \cite{VB12} for details).

The performances of the imputation methods (with or without variable selection) are assessed by considering the bias, the empirical variance and the mean squared error on the mean of each variable. Results are provided in Tables \ref{decatlon} and \ref{wine}.

\begin{table}[ht]
\centering
\caption{Decathlon data set: bias, empirical variance and mean squared error for the MI estimators of the mean for each variable. Two MI methods are considered: by considering all explanatory variables in the conditional imputation models (all) or by specifying models using variable selection (select).\label{decatlon}}
\begin{tabular}{rrrrrrr}
 \hline
  &\multicolumn{2}{c}{Bias}&\multicolumn{2}{c}{Empirical variance}&\multicolumn{2}{c}{Mean squared error}\\ \hline
 & select & all & select & all & select & all \\
  \hline

V1 & 0.008 & 0.012 & 0.010 & 0.011 & 0.010 & 0.011 \\
  V2 & 0.011 & 0.015 & 0.011 & 0.011 & 0.011 & 0.011 \\
  V3 & 0.001 & 0.007 & 0.012 & 0.009 & 0.012 & 0.009 \\
  V4 & 0.018 & 0.005 & 0.014 & 0.016 & 0.014 & 0.016 \\
  V5 & 0.014 & 0.003 & 0.015 & 0.010 & 0.016 & 0.010 \\
  V6 & 0.004 & 0.017 & 0.012 & 0.013 & 0.012 & 0.013 \\
  V7 & 0.008 & 0.003 & 0.012 & 0.012 & 0.012 & 0.012 \\
  V8 & 0.016 & 0.025 & 0.018 & 0.019 & 0.018 & 0.019 \\
  V9 & 0.002 & 0.012 & 0.013 & 0.015 & 0.013 & 0.015 \\
  V10 & 0.001 & 0.005 & 0.018 & 0.017 & 0.018 & 0.017 \\
  V11 & 0.008 & 0.002 & 0.012 & 0.003 & 0.012 & 0.003 \\
   \hline
\end{tabular}
\end{table}

\begin{table}[ht]
\centering
\caption{Wine data set: bias, empirical variance and mean squared error for the MI estimators of the mean for each variable. Two MI methods are considered: by considering all explanatory variables in the conditional imputation models (all) or by specifying models using variable selection (select).\label{wine}}
\begin{tabular}{rrrrrrr}
  \hline
  &\multicolumn{2}{c}{Bias}&\multicolumn{2}{c}{Empirical variance}&\multicolumn{2}{c}{Mean squared error}\\ \hline
 & select & all & select & all & select & all \\
  \hline
V1 & 0.00 & 1.89 & 0.02 & 473.80 & 0.02 & 477.37 \\
  V2 & 0.01 & 0.08 & 0.02 & 71.35 & 0.02 & 71.35 \\
  V3 & 0.00 & 0.79 & 0.02 & 96.09 & 0.02 & 96.71 \\
  V4 & 0.03 & 0.04 & 0.02 & 406.97 & 0.03 & 406.97 \\
  V5 & 0.02 & 4.90 & 0.02 & 1470.50 & 0.02 & 1494.54 \\
  V6 & 0.00 & 1.67 & 0.01 & 142.36 & 0.01 & 145.17 \\
  V7 & 0.01 & 2.75 & 0.01 & 6994.86 & 0.01 & 7002.41 \\
  V8 & 0.02 & 4.48 & 0.01 & 1517.93 & 0.01 & 1538.01 \\
  V9 & 0.02 & 0.04 & 0.02 & 56.21 & 0.02 & 56.21 \\
  V10 & 0.02 & 1.88 & 0.02 & 511.96 & 0.02 & 515.48 \\
  V11 & 0.01 & 2.63 & 0.02 & 917.00 & 0.02 & 923.92 \\
  V12 & 0.02 & 0.17 & 0.03 & 672.40 & 0.03 & 672.43 \\
  V13 & 0.03 & 1.01 & 0.03 & 607.86 & 0.03 & 608.87 \\
  V14 & 0.03 & 2.57 & 0.03 & 1470.79 & 0.03 & 1477.39 \\
  V15 & 0.01 & 0.62 & 0.02 & 461.67 & 0.02 & 462.05 \\
  V16 & 0.00 & 0.54 & 0.02 & 55.23 & 0.02 & 55.52 \\
  V17 & 0.01 & 0.68 & 0.01 & 276.85 & 0.01 & 277.32 \\
  V18 & 0.02 & 0.52 & 0.02 & 66.19 & 0.02 & 66.46 \\
  V19 & 0.01 & 0.81 & 0.02 & 58.83 & 0.02 & 59.49 \\
  V20 & 0.00 & 2.98 & 0.04 & 684.04 & 0.04 & 692.92 \\
  V21 & 0.03 & 0.60 & 0.02 & 139.83 & 0.02 & 140.20 \\
  V22 & 0.01 & 6.48 & 0.02 & 4496.05 & 0.02 & 4538.03 \\
  V23 & 0.02 & 0.02 & 0.01 & 114.61 & 0.01 & 114.61 \\
  V24 & 0.01 & 1.83 & 0.01 & 220.76 & 0.01 & 224.11 \\
  V25 & 0.02 & 1.50 & 0.03 & 551.10 & 0.03 & 553.34 \\
  V26 & 0.03 & 0.11 & 0.02 & 29.99 & 0.02 & 30.00 \\
  V27 & 0.01 & 40.48 & 0.01 & 233135.03 & 0.01 & 234774.03 \\
  V28 & 0.00 & 2.93 & 0.01 & 2254.54 & 0.01 & 2263.14 \\
  V29 & 0.02 & 0.51 & 0.01 & 109.07 & 0.01 & 109.33 \\
   \hline
\end{tabular}
\end{table}

Results on the wine data set highlight specifying conditional models by the proposed algorithm can drastically decreases the bias and the variance of the MI estimators. The benefit to use variable selection on the decathlon data set is less clear, but performances remain similar.

\section{Discussion}
High-dimensional data as well as missing data are two of the main challenges for applied statistician at the digital era.
In this article we proposed an algorithm for variable selection in the framework of linear models. This algorithm improves the performances of many selection methods (in terms of true positive and false positive rates) and provides a measure of importance for the explanatory variables. Furthermore, it allows handling missing values (MAR or MCAR) and/or high-dimensional settings for any variable selection method. From a practical point of view, the method has the advantage to allow parallel calculation, solving some potential calculation time issues. In addition, its parameters can be easily tuned: the number of iterations $\Nbiter$ can be checked by inspecting stability of proportions of selection, while the number of variables drawn ($\Nbvardraw$) can be chosen a priori (since the method is robust to this parameter) and the threshold $r$ can be chosen by cross-validation. Among its applications, we showed it can be effectively applied for model specification in sequential imputation by offering a valuable way to automatically tune conditional imputation models.

Various extensions of the algorithm can be proposed. First, we assumed Gaussian explanatory
variables, but this Gaussian assumption can be easily raised. For achieving this goal, multiple imputation by chained equations is appealing since it allows the use of semi-parametric imputation models \cite{Morris14,Gaffert18}. Second, we assume a Gaussian distribution for the response variable given explanatory variables. The algorithm can be easily adapted in the case of Generalized Linear Models (GLM) or mixed models by using suitable selection variables methods (see e.g. \cite{GegoutPetit20}), but additional statistical work has to be done to tune the parameters. Third, we did not explore the specific case of data missing not at random, but the algorithm could be adapted to accounting for such mechanisms by using suitable imputation method \cite[e.g.]{Galimard16}.

Refinements of the algorithm could also be possible. In particular, accounting for the variation around $r_\nbvar$ in the threshold could be quite easy. This could accelerate the algorithm by limiting the number of iterations without ignoring the Monte Carlo error. Another refinement would be to use multiple imputation instead of stochastic single imputation, but this approach remains tricky from a computational point of view: first, single imputation only requires an Expectation-Maximization (EM) algorithm \cite{Dempster77}, while multiple imputation by Gaussian model \cite{Schafer97} requires a Data-Augmentation algorithm \cite{Tanner87} which should be previously initialised by an EM algorithm. Second, the variable selection should be performed on each imputed data set, while it is applied only one time with single imputation. Third, results required to be pooled, while single imputation does not require any pooling step.

Outliers is also a classical problem in data analysis. While robust estimates can be considered (see~\cite{Alfons13} for example), it is also possible to remove them by replacing them with missing values. Therefore, this algorithm could be advocated to handle outliers in variable selection.

Moreover, in this article we fixed a threshold to include (or not) a variable in the model for a given instance, but we could also aggregate the probabilities (under the null hypothesis) that $\beta_\nbvar=0$ . A natural aggregation over all instances is given by the empirical mean, that can be seen as the mean of the estimates of $\mathbb{P}(\beta_\nbvar=0)$. Then, for each variable, this mean would be thresholded, as proposed.  

Finally, we focused on variable selection, but one may notice that each instance gives estimates of $\beta$ and we can also aggregate these estimates. However, such an extension is not straightforward since estimates are generally biased on all instances. Further research on aggregation of those biased estimates could lead to the development of a robust estimator of regression coefficients in a high-dimensional setting with missing values.

\clearpage
\bibliographystyle{unsrt}
\bibliography{biblio}
\newpage
\appendix

\section{Low dimensional data}
~~
\begin{figure}[h!]
\center
\includegraphics[scale=.5]{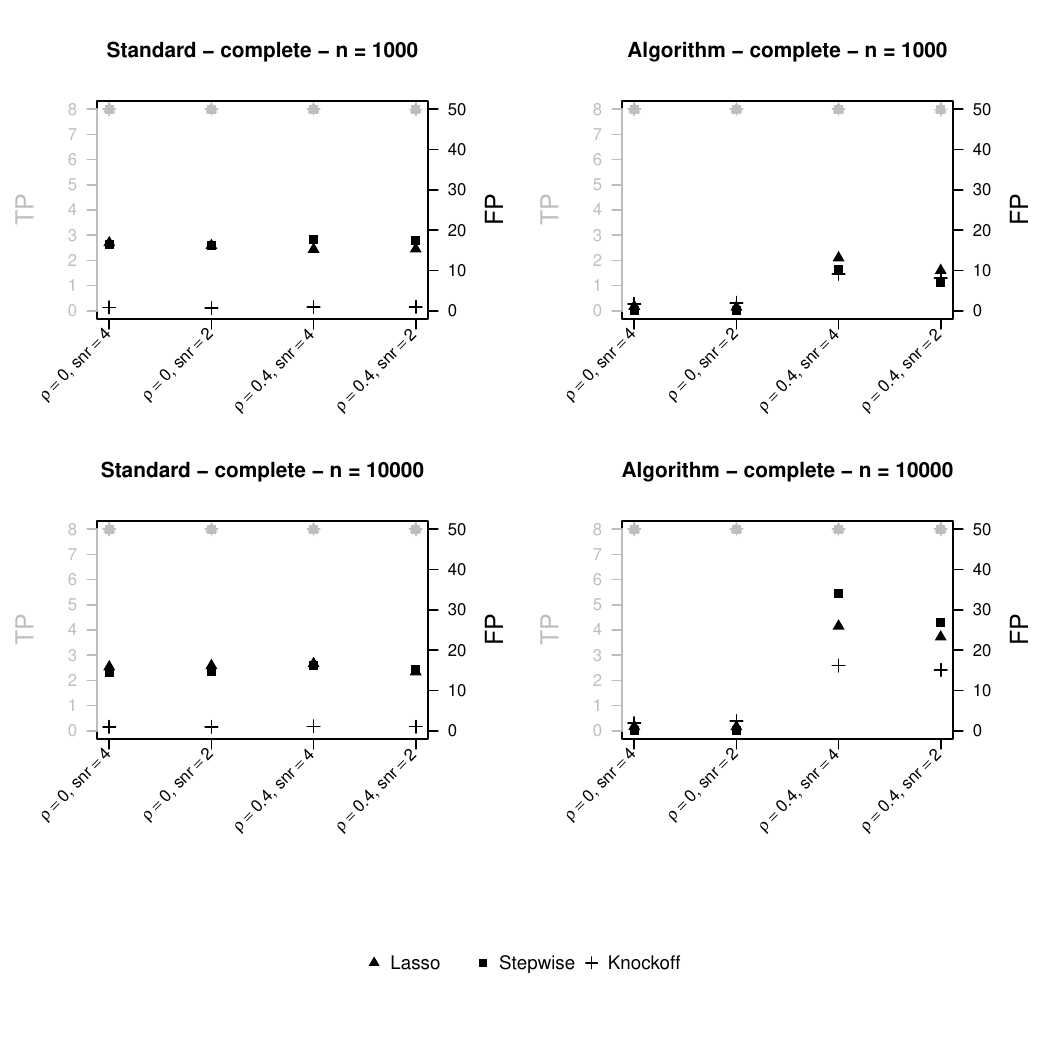}
\caption{Assessment of the ensemble method according to the number of individuals and the data structure. Figure reports results when $\Nbvar=100$ for $\Nbind=1000$ (on the top) or $\Nbind=10 000$ individuals (on the bottom). Assessment is based on data sets varying by the correlation between covariates ($\rho$), the signal to noise ratio ($snr$). For a given configuration, $\Nbsim$ data sets are generated. Three variable selection methods (Knockoff, Lasso and Stepwise) are compared when they are directly applied on the data set (standard) or when they are iteratively applied on subsets of variables (algorithm). Performances of the selection procedure are assessed by: the mean number of true positives (in grey) and the mean number of false positives (in black) over the $\Nbsim$ data sets.\label{figure_pt_dim_plusind}}
\end{figure}

\begin{table}[h!]
\centering
\footnotesize
\begin{tabular}{lrllrrrrrr}
  \hline
  &&&&\multicolumn{3}{c}{Algorithm}&\multicolumn{3}{c}{Standard} \\ \hline

  $\rho$ & snr & mech &method& TP&FN&FP&TP&FN&FP \\
  \hline
0 & 2 & MCAR & Knockoff & 6.75 & 1.25 & 1.47 &  &  &  \\
  0 & 2 & MAR & Knockoff & 6.78 & 1.22 & 2.18 & 0.18 & 7.82 & 0.44 \\
  0 & 4 & MCAR & Knockoff & 7.43 & 0.57 & 1.14 &  &  &  \\
  0 & 4 & MAR & Knockoff & 7.24 & 0.76 & 1.90 & 0.11 & 7.89 & 0.49 \\
  0.4 & 2 & MCAR & Knockoff & 4.32 & 3.68 & 1.92 &  &  &  \\
  0.4 & 2 & MAR & Knockoff & 4.79 & 3.21 & 2.62 & 0.07 & 7.93 & 0.47 \\
  0.4 & 4 & MCAR & Knockoff & 6.15 & 1.85 & 2.45 &  &  &  \\
  0.4 & 4 & MAR & Knockoff & 6.35 & 1.65 & 2.88 & 0.04 & 7.96 & 0.52 \\
  0 & 2 & MCAR & Lasso & 7.09 & 0.91 & 1.74 &  &  &  \\
  0 & 2 & MAR & Lasso & 7.21 & 0.79 & 3.13 & 0.45 & 7.55 & 2.62 \\
  0 & 4 & MCAR & Lasso & 7.75 & 0.25 & 1.59 &  &  &  \\
  0 & 4 & MAR & Lasso & 7.64 & 0.36 & 3.26 & 0.58 & 7.42 & 3.49 \\
  0.4 & 2 & MCAR & Lasso & 5.40 & 2.60 & 3.14 &  &  &  \\
  0.4 & 2 & MAR & Lasso & 5.77 & 2.23 & 4.17 & 0.21 & 7.79 & 1.83 \\
  0.4 & 4 & MCAR & Lasso & 6.85 & 1.15 & 4.21 &  &  &  \\
  0.4 & 4 & MAR & Lasso & 7.01 & 0.99 & 4.96 & 0.18 & 7.82 & 2.10 \\
  0 & 2 & MCAR & Stepwise & 6.11 & 1.89 & 0.46 &  &  &  \\
  0 & 2 & MAR & Stepwise & 6.48 & 1.52 & 1.07 &  &  &  \\
  0 & 4 & MCAR & Stepwise & 7.14 & 0.86 & 0.31 &  &  &  \\
  0 & 4 & MAR & Stepwise & 7.18 & 0.82 & 1.25 &  &  &  \\
  0.4 & 2 & MCAR & Stepwise & 3.88 & 4.12 & 1.07 &  &  &  \\
  0.4 & 2 & MAR & Stepwise & 4.42 & 3.58 & 1.76 &  &  &  \\
  0.4 & 4 & MCAR & Stepwise & 5.91 & 2.09 & 1.77 &  &  &  \\
  0.4 & 4 & MAR & Stepwise & 5.95 & 2.05 & 2.44 &  &  &  \\    \hline
\end{tabular}
\caption{Low dimensional setting with missing values: performances of three variable selection methods (Knockoff, Lasso and Stepwise) when they are iteratively applied on imputed subsets of variables (Algorithm) or when they are applied on the complete individuals of the data set (Standard). Missing values are related to failure because of a too low number of complete-cases. Data sets varying by the correlation between covariates ($\rho$) and the signal to noise ratio ($snr$). For a given configuration, $\Nbsim$ data sets are generated and performances of the selection procedure are assessed by: the mean number of true positives (TP), the mean number of false negatives (FN) and the mean number of false positives (FP) (the number of real positives is 8 and the number of real negatives is 92). \label{table_ptdim_incomplet}}

\end{table}\begin{figure}
\begin{center}
\includegraphics[width=10cm,height=8cm]{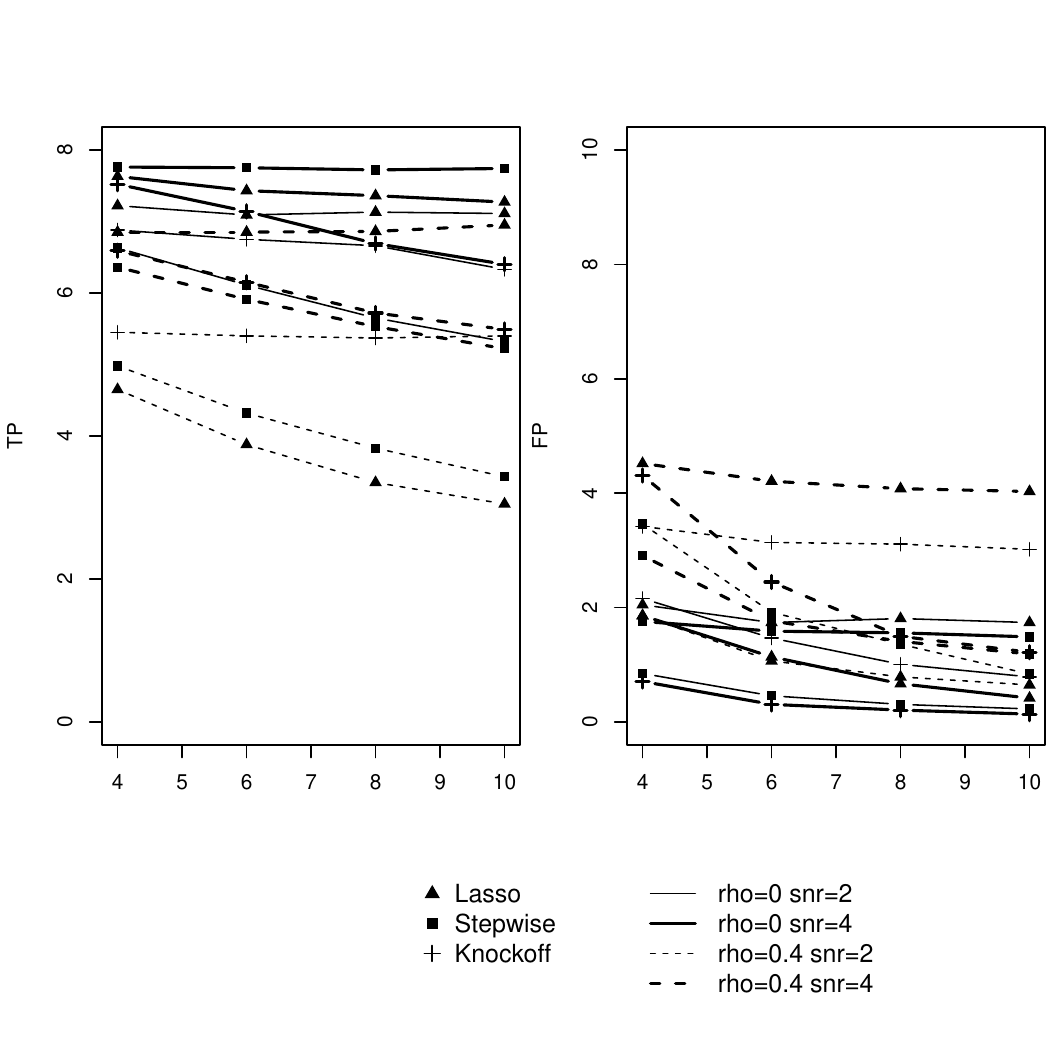}
\caption{Influence of $\Nbvardraw$ in the low dimensional setting with values missing completely at random: true positive rate (on the left) and false positive rate (on the right) according to the number of variables sampled in the algorithm ($\Nbvardraw$) for the 4 configurations varying by the signal to noise ratio (snr), the correlation between covariates ($\rho$). Three variable selection methods are reported (lasso, stepwise and knockoff).}\label{fig_k_incomplet_mcar}
\end{center}
\end{figure}\begin{figure}
\begin{center}
\includegraphics[width=10cm,height=8cm]{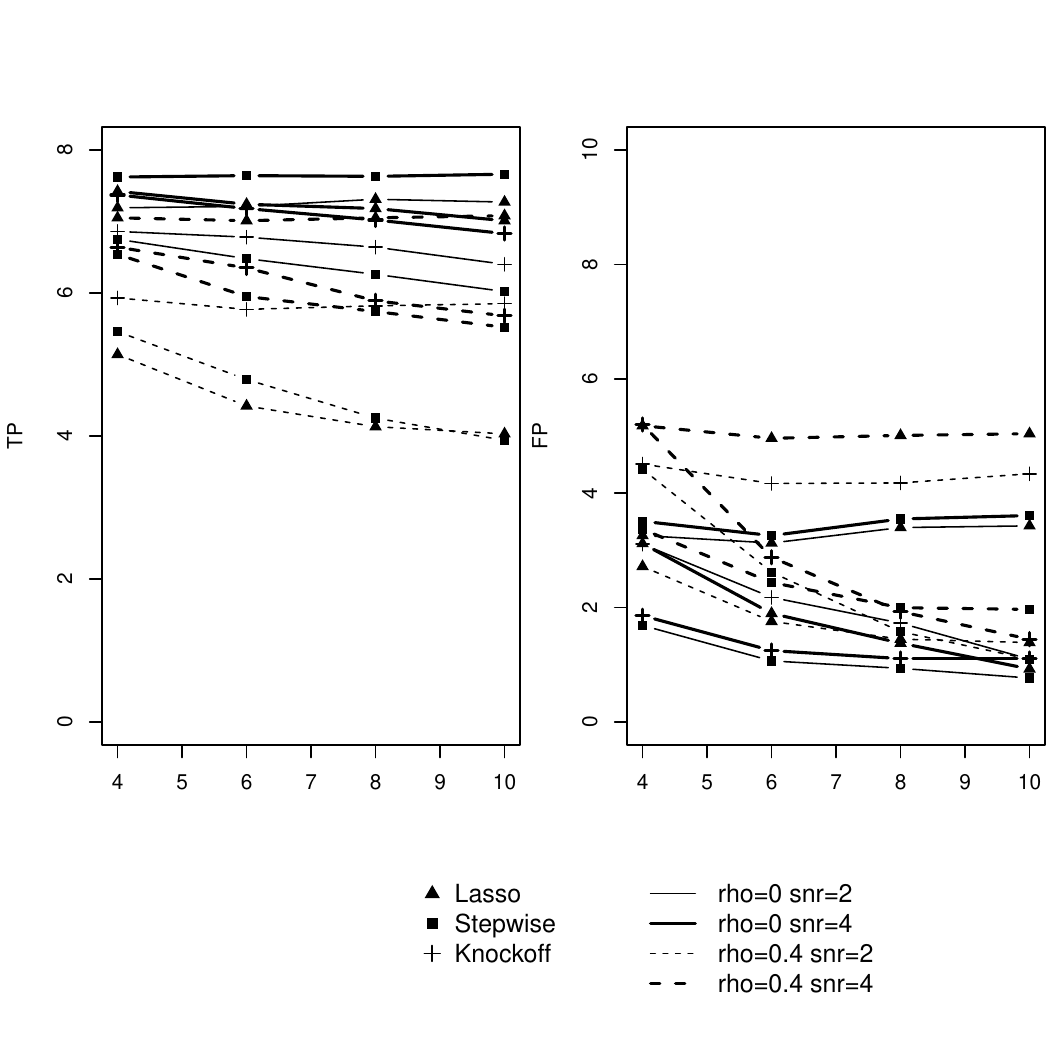}
\caption{Influence of $\Nbvardraw$ in the low dimensional setting with values missing at random: true positive rate (on the left) and false positive rate (on the right) according to the number of variables sampled in the algorithm ($\Nbvardraw$) for the 4 configurations varying by the signal to noise ratio (snr), the correlation between covariates ($\rho$). Three variable selection methods are reported (lasso, stepwise and knockoff).}\label{fig_k_incomplet_mar}
\end{center}
\end{figure}
\clearpage

\section{High dimensional data}
\begin{table}[h]
\centering
\footnotesize
\begin{tabular}{lrllrrrrrr}
  \hline
    &&&&\multicolumn{3}{c}{Algorithm}&\multicolumn{3}{c}{Standard}\\ \hline
$\rho$ & snr & mech & method & TP & FN & FP & TP & FN & FP \\
  \hline
0 & 2 & MCAR & Knockoff & 6.58 & 1.42 & 3.57 &  &  &  \\
  0 & 2 & MAR & Knockoff & 6.32 & 1.68 & 6.22 & 0.02 & 7.98 & 0.77 \\
  0 & 4 & MCAR & Knockoff & 7.20 & 0.80 & 4.21 &  &  &  \\
  0 & 4 & MAR & Knockoff & 7.19 & 0.81 & 6.69 & 0.04 & 7.96 & 0.52 \\
  0.4 & 2 & MCAR & Knockoff & 4.04 & 3.96 & 4.76 &  &  &  \\
  0.4 & 2 & MAR & Knockoff & 4.32 & 3.68 & 6.52 & 0.03 & 7.97 & 0.53 \\
  0.4 & 4 & MCAR & Knockoff & 5.51 & 2.49 & 4.75 &  &  &  \\
  0.4 & 4 & MAR & Knockoff & 5.78 & 2.22 & 6.19 & 0.00 & 8.00 & 0.47 \\
  0 & 2 & MCAR & Lasso & 7.11 & 0.89 & 4.39 &  &  &  \\
  0 & 2 & MAR & Lasso & 6.96 & 1.04 & 8.63 & 0.14 & 7.86 & 3.07 \\
  0 & 4 & MCAR & Lasso & 7.58 & 0.42 & 4.02 &  &  &  \\
  0 & 4 & MAR & Lasso & 7.45 & 0.55 & 8.48 & 0.08 & 7.92 & 2.66 \\
  0.4 & 2 & MCAR & Lasso & 4.42 & 3.58 & 5.22 &  &  &  \\
  0.4 & 2 & MAR & Lasso & 4.77 & 3.23 & 7.33 & 0.03 & 7.97 & 1.83 \\
  0.4 & 4 & MCAR & Lasso & 6.33 & 1.67 & 5.59 &  &  &  \\
  0.4 & 4 & MAR & Lasso & 6.40 & 1.60 & 7.53 & 0.05 & 7.95 & 1.88 \\
  0 & 2 & MCAR & Stepwise & 6.01 & 1.99 & 0.71 &  &  &  \\
  0 & 2 & MAR & Stepwise & 6.18 & 1.82 & 2.92 &  &  &  \\
  0 & 4 & MCAR & Stepwise & 6.77 & 1.23 & 0.80 &  &  &  \\
  0 & 4 & MAR & Stepwise & 6.79 & 1.21 & 2.79 &  &  &  \\
  0.4 & 2 & MCAR & Stepwise & 2.62 & 5.38 & 1.08 &  &  &  \\
  0.4 & 2 & MAR & Stepwise & 3.22 & 4.78 & 2.13 &  &  &  \\
  0.4 & 4 & MCAR & Stepwise & 4.65 & 3.35 & 1.36 &  &  &  \\
  0.4 & 4 & MAR & Stepwise & 5.04 & 2.96 & 2.05 &  &  &  \\
   \hline
\end{tabular}
\caption{High-dimensional setting with missing values: performances of three variable selection methods (Knockoff, Lasso and Stepwise) when they are iteratively applied on imputed subsets of variables (Algorithm) or when they are applied on the complete individuals of the data set (Standard). Missing values are related to failure because of a too low number of complete-cases. Data sets varying by the correlation between covariates ($\rho$) and the signal to noise ratio ($snr$). For a given configuration, $\Nbsim$ data sets are generated and performances of the selection procedure are assessed by: the mean number of true positives (TP), the mean number of false negatives (FN) and the mean number of false positives (FP) (the number of real positives is 8 and the number of real negatives is 92). \label{table_gddim_incomplet}}
\end{table}

\begin{figure}
\begin{center}
\includegraphics[width=10cm,height=8cm]{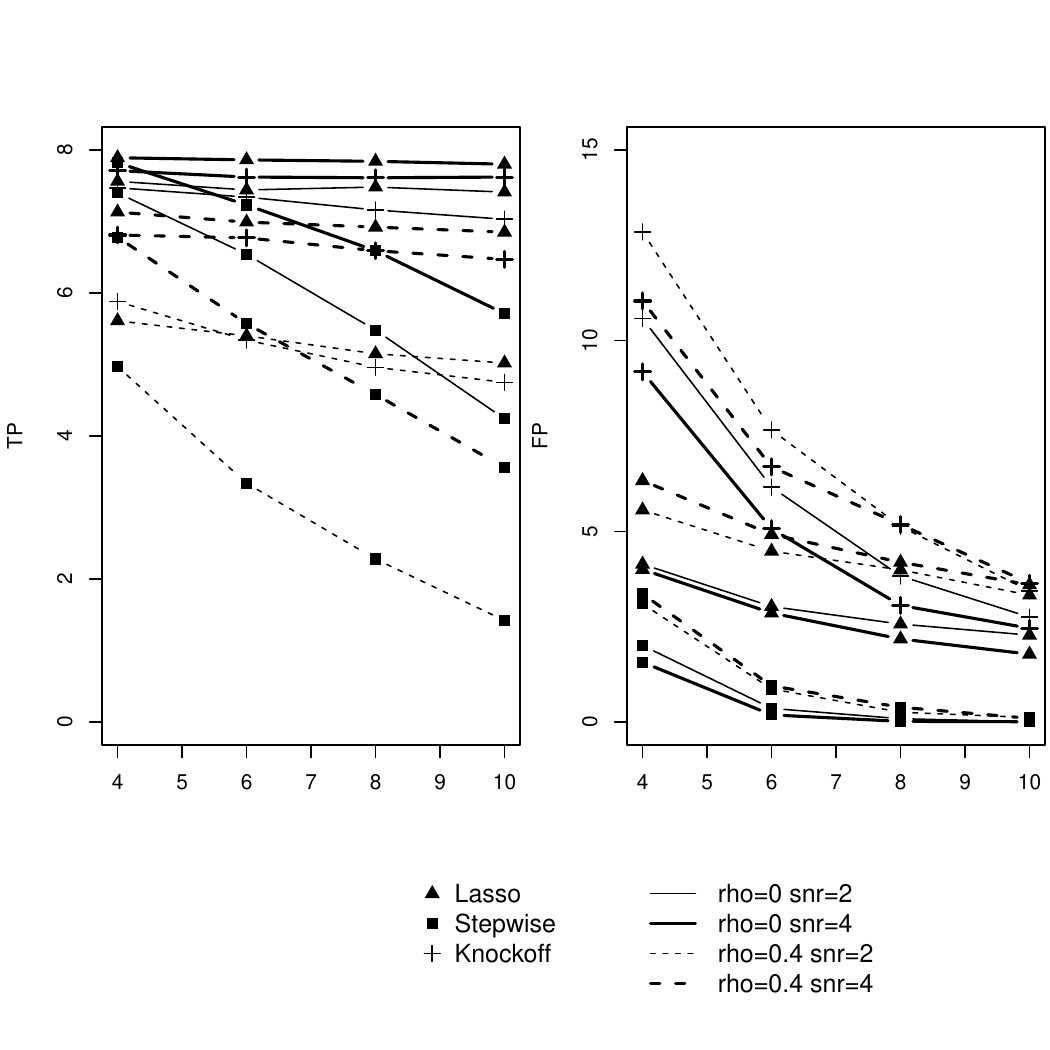}
\caption{Influence of $\Nbvardraw$ in the high-dimensional setting without missing values: true positive rate (on the left) and false positive rate (on the right) according to the number of variables sampled in the algorithm ($\Nbvardraw$) for the 4 configurations varying by the signal to noise ratio (snr), the correlation between covariates ($\rho$). Three variable selection methods are reported (lasso, stepwise and knockoff).}\label{fig_k_complet_gddim}
\end{center}
\end{figure}

\begin{figure}
\begin{center}
\includegraphics[width=10cm,height=8cm]{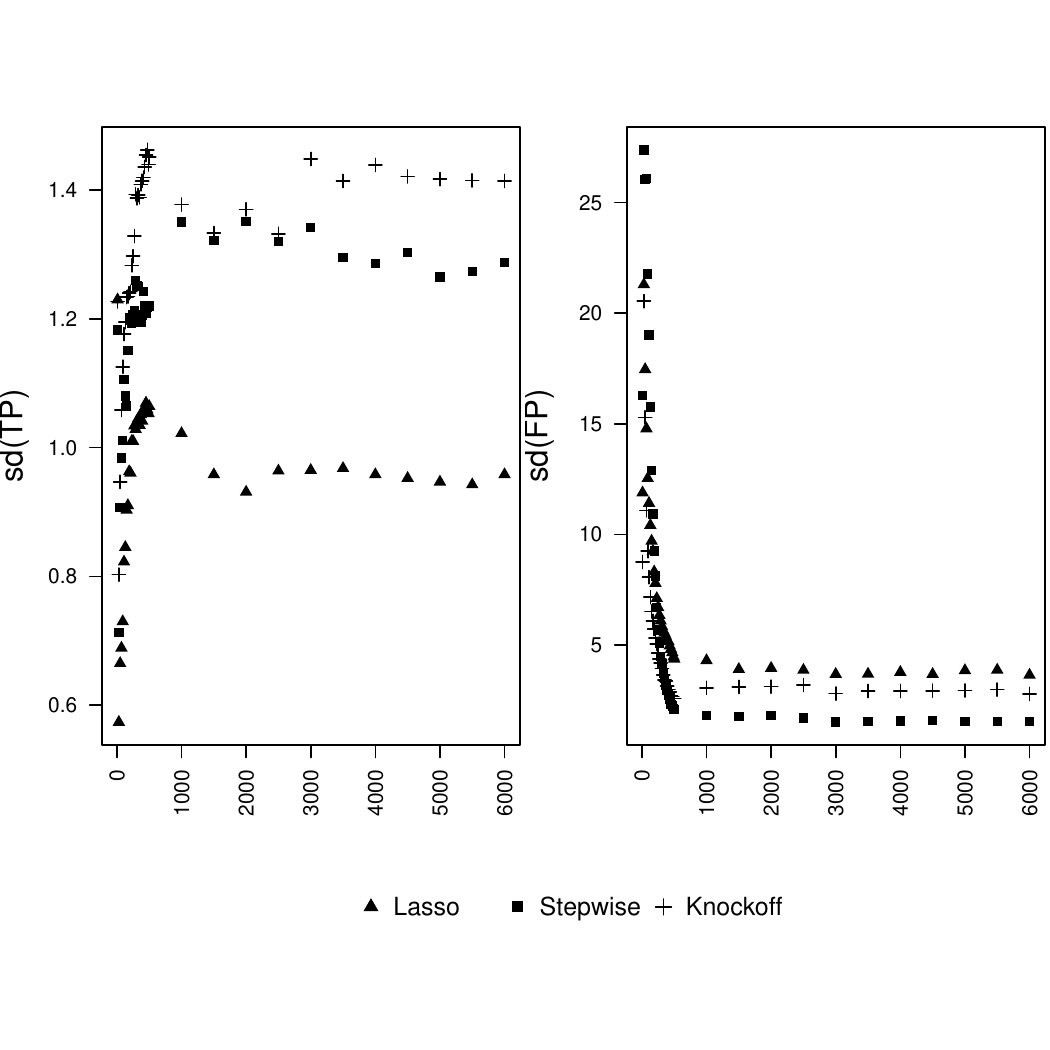}
\caption{Influence of $\Nbiter$ for high-dimensional setting with missing values: standard deviation of the true positive rate over the 100 generated data sets (on the left) and false positive rate (on the right) according to the number of iterations of the algorithm ($\Nbiter$) for the configuration with signal to noise ratio equal to 2, null correlation between covariates ($\rho$) and values missing at random. 3 variable selection methods are reported (lasso, stepwise and knockoff).}\label{fig_B_incomplet}
\end{center}
\end{figure}
\end{document}